\documentclass[10pt,preprintnumbers,nofootinbib]{revtex4-2}%
\usepackage[latin9]{inputenc}
\usepackage{amsmath}
\usepackage{amssymb}
\usepackage{float}
\usepackage{dcolumn}
\usepackage{graphicx}
\usepackage[colorlinks=true, pdfstartview=FitV, linkcolor=blue, citecolor=red, urlcolor=magenta]{hyperref}
\usepackage{amsfonts}
\usepackage{subfigure}
\usepackage{cleveref}
\usepackage{listings}
\usepackage{xcolor}
\usepackage{array}
\usepackage{url}
\usepackage{color}%
\usepackage{tikz}
\usepackage{pgfplots}
\usepackage{todonotes}
\usepackage{multirow}
\usepackage[outdir=./Figures]{epstopdf}

\newcommand{\oo}{\mathring}

\usepackage{booktabs}

\newcommand{\vp}{\varphi}
\usepackage{xcolor}

\graphicspath{{Figures/}}

\begin{document}

\begin{flushright}
\texttt{ZTF-EP-25-08}

\texttt{RBI-ThPhys-2025-42}
\end{flushright}

\vspace{20pt}

\title{Noncommutative Regge-Wheeler potential: some nonperturbative results}

\author{Nikola Herceg}
\email{nherceg@irb.hr}
\author{Tajron Juri\'c}
\email{tjuric@irb.hr}
\author{A. Naveena Kumara }
\email{nathith@irb.hr}
\author{Andjelo Samsarov}
\email{asamsarov@irb.hr}
\affiliation{Rudjer Bo\v{s}kovi\'c Institute, Bijeni\v cka c. 54, HR-10002 Zagreb, Croatia}

\author{Ivica Smoli\'c}
\email{ismolic@phy.hr}
\affiliation{Department of Physics, Faculty of Science, University of Zagreb, Bijeni\v cka cesta 32, 10000 Zagreb, Croatia}
\date{\today}

\begin{abstract}
	We study the gravitational perturbation theory of black holes in noncommutative spacetimes with noncommutativity of the type
$[t\stackrel{\star}{,} r] = i a \alpha A(r)$ and $[\varphi \stackrel{\star}{,} r] = i a \beta A(r)$ for arbitrary $A(r)$, which includes several Moyal-type spaces and also the $\kappa$-Minkowski space.
The main result of this paper is an analytical expression for the effective potential of the axial perturbation modes, valid to all orders in the noncommutativity parameter.
This is achieved by evaluating the $\star$-products using translations in the radial direction, i.e., Bopp shift. We comment on various regimes, such as Planck-scale black holes, where the noncommutativity length scale is of the same order of magnitude as the black hole horizon.\\ \\
\end{abstract}

\maketitle

\section{Introduction}

The majority of theoretical approaches to quantum gravity, such as string theory~\cite{Polchinski:1998rq, Mukhi:2011zz} and loop quantum gravity~\cite{Rovelli:2004tv, Ashtekar:2021kfp}, anticipate that certain features of novel and distinctive character will begin to emerge near the Planck length scale \( l_{\text{Pl}} \sim \theta\), where \( \theta \) is the deformation parameter of the theory. Some of the most notable features of this kind involve the violation of Lorentz symmetry~\cite{Vasileiou:2015wja}, gravitationally mediated entanglement~\cite{Lindner:2004bw}, and decoherence induced by fluctuations in the spacetime foam~\cite{Oniga:2015lro,Oniga:2017pyq,Oniga:2016ubb}. These features are mainly sought in light from gamma-ray bursts, astrophysical and atmospheric neutrinos~\cite{IceCube:2021tdn,ICECUBE:2023gdv}, and the imprints of quantum gravitational effects on the polarization of the cosmic microwave background radiation~\cite{Calmet:2019tur}, to name a few. All these phenomena presumably appear at sufficiently high energies. However, when the low-energy limit of these more fundamental theories of gravity is taken, resulting in effective theory descriptions, the aforementioned peculiar features are expected to survive the limit. They remain as the only testimonies to quantum gravity, acting as remnants that have persisted and continue to manifest at lower energy scales.\\

In order to pinpoint and describe these features of quantum gravity that are supposed to survive the low-energy limit, researchers regularly use certain effective theoretical frameworks based on the \( \theta \)-expansion method, which involves a perturbative expansion in terms of the Planck scale parameter \( \theta \). However, although the \( \theta \)-expansion method works well in model building, essential information of a nonperturbative origin is lost because of the cutoff at a finite order in \( \theta \). For example, in the area of noncommutative phenomenology, it is well-known that the Moyal-Weyl $\star$-product in noncommutative field theories~\cite{Minwalla:1999px,Hayakawa:1999yt,Hayakawa:1999zf,Matusis:2000jf} gives rise to a nontrivial phase factor. This is revealed by considering the corresponding Fourier modes when two functions are multiplied together. While this phase factor regulates the ultraviolet divergence in the one-loop two-point function of certain noncommutative field theories at the level of one-loop calculations, it simultaneously introduces an infrared-divergent term of the form \( 1/{(\theta p^2)} \), where \( p \) is the external momentum. This is the origin of the famous UV/IR mixing~\cite{Minwalla:1999px}.\\

The nontrivial phase factor appears only when all orders of \( \theta \) in the $\star$-product are taken into account and summed over. Consequently, the feature of UV/IR mixing does not manifest itself when noncommutative gauge theories are studied using the perturbative \( \theta \)-expansion~\cite{Minwalla:1999px}.
Noncommutative gauge theories can, however, be formulated in other ways. For example, by introducing certain generalized $\star$-products~\cite{Jurco:2001rq,Mehen:2000vs}, it is possible to carry out a perturbative expansion only with respect to the gauge coupling constant, much like in ordinary gauge field theory. In this approach, each individual term in the expansion already includes all orders of \( \theta \). This expansion allows for handling all orders of \( \theta \) simultaneously at every step of the calculation, thereby facilitating the computation of nonperturbative results. Characteristic examples of this approach, particularly when applied to noncommutative (NC) gauge field theories, include computing the fermion one-loop correction to the photon two-point function in certain models of noncommutative electrodynamics~\cite{Schupp:2008fs}, computing one-loop 1PI (one-particle irreducible) contributions to all the propagators of noncommutative super Yang-Mills \( U(1) \) theories~\cite{Martin:2016zon}, and studying various aspects of NC photon-neutrino phenomenology~\cite{Horvat:2011iv}. These aspects include the scattering of ultra-high-energy cosmic ray neutrinos on nuclei~\cite{Horvat:2011iv}, plasmon decay~\cite{Trampetic:2002eb}, big bang nucleosynthesis~\cite{Horvat:2009cm}, and self-energies in deformed spacetimes~\cite{Horvat:2013rga}. In all these cases, the analysis was carried out using the \( \theta \)-exact covariant noncommutative field theory framework, resulting in the reappearance of UV/IR mixing. This demonstrates that the previously observed absence of UV/IR mixing within the perturbative \( \theta \)-expansion approach to noncommutative gauge theory was merely a technical artifact of the method, rather than a genuine characteristic feature of the theory itself.\\

Generally speaking, when extracting genuine effects of spacetime distortion in noncommutative phenomenology and avoiding interference with dominant background contributions from familiar processes, attention has shifted toward processes that are known to be suppressed in Standard Model settings. To prevent difficulties in interpreting results that are hard to distinguish from dominant underlying signals, the focus is on interactions that, in the absence of noncommutativity, are either forbidden or unlikely to occur. Any nonzero contribution found in these processes would then unambiguously indicate an NC effect. A common example in NC phenomenology is the tree-level coupling of neutrinos with photons, which does not exist in the Standard Model. However, the NC setting (noncommutative field theories on noncommutative spaces) introduces new and different interaction channels for neutrinos and photons. This allows for their tree-level coupling, resulting in nonvanishing tree-level cross sections and decay rates in processes involving photons and neutrinos, such as the transverse plasmon decay into a neutrino-antineutrino pair~\cite{Schupp:2002up}.\\

Parallel to the noncommutative phenomenology related to the Standard Model, relatively small but nevertheless sound effort has also been directed toward phenomenology related to  noncommutative gravity ~\cite{Kobakhidze:2016cqh,Jenks:2020gbt,Campos:2021sff,Zhao:2023uam,Yan:2020hga,Gupta:2017lwk,Gupta:2015uga,Ciric:2017rnf,DimitrijevicCiric:2019hqq} and other frameworks encompassing the quantum nature of spacetime like loop quantum gravity~\cite{Cruz:2015bcj}.
In noncommutative gravity phenomenology, the main emphasis is placed on the notion of black hole quasinormal modes.
Black hole quasinormal modes are important for many reasons, among them the issue  of black hole stability and the possibility of providing a more direct contact with black holes through the experimental observation of gravitational waves.
At the same time, the discovery of gravitational waves~\cite{LIGOScientific:2016aoc, LIGOScientific:2016sjg, LIGOScientific:2017vwq, Castelvecchi2016} has made the doors wide open to new perspectives in the investigation of the spacetime structure, including its quantum spacetime characteristics.  
The study of the effects of the quantum structure of spacetime on gravitational wave formation thus appears to be a quite natural endeavor.\\

First steps in this direction were taken in~\cite{Herceg:2023zlk, Herceg:2023pmc}, where the study of the axial gravitational perturbations in the NC Schwarzschild background is made. Later on, the derivation of polar modes in the same background has led to the discovery of isospectral breaking in the polar and axial modes, a phenomenon that does not appear in the absence of noncommutativity. When noncommutativity is introduced, the breaking of isospectrality between axial and polar modes of the Schwarzschild black hole becomes evident~\cite{Herceg:2024vwc}.\\

In this paper, we use the method of \( \theta \)-exact calculus to obtain nonperturbative equations of motion of axial gravitational waves in noncommutative Schwarzschild spacetime. We focus on the noncommutativity occurring in the radial direction with arbitrary radial dependence, such as\footnote{From now on, we will use $a$ for the noncommutativity parameter.} $[r,t] = i a A(r)$.
For some choices of $A(r)$, the effective potential governing the gravitational wave differs significantly from the classical single-peak Regge-Wheeler potential.
The deviation becomes even more prominent as the Schwarzschild radius approaches the Planck scale, turning the single-peak structure into a potential barrier in the case of $\kappa$-Minkowski. \\

The paper is organized as follows: Section \ref{sec2} introduces a special class of semipseudo-Killing twists that generate Bopp shifts and demonstrates how these Bopp shifts can be used to evaluate $\star$-products nonperturbatively. In Section \ref{sec3}, this Bopp shift-based calculus is applied to linear gravitational perturbation theory in noncommutative spacetime. Regarding the noncommutativity parameter, this calculus allows us to perform computations in a completely nonperturbative manner, with the gravitational perturbation \( h \) itself remaining linear. The results of this calculation are then used in Section \ref{sec4} to analyze axial mode potentials for gravitational perturbations that arise from different choices of $\star$-products (i.e., functions $A(r)$). We conclude with some final remarks in Section \ref{conclusion}.


\section{Twist, $\star$-product and some useful identities} \label{sec2}
In this section, we will introduce the class of twists that will be used in the rest of the paper and demonstrate how $\star$-products are evaluated nonperturbatively for certain classes of functions.
Fixing a specific type of noncommutativity is necessary since the evaluation of the $\star$-product is influenced by the symmetry of the background spacetime.  
A formal route of introducing the $\star$-product, which replaces the usual pointwise product of smooth functions, goes by twisting the Hopf algebra of diffeomorphisms by a twist element $\mathcal{F}$.
We study the class of Moyal twists consisting of two commuting vector fields.
More precisely, we employ a semipseudo-Killing twist of the form
\begin{equation} \label{radialtwist}
	\mathcal F = \exp \frac{- i a}{2} (K \otimes X - X \otimes K),
\end{equation}
where 
\begin{equation}
K = \alpha \partial_t + \beta \partial_\varphi, \quad X = A(r) \partial_r,
\end{equation}
are respectively a Killing field of the Schwarzschild metric and an arbitrary radially directed vector field.
The term semipseudo-Killing means that the twist isn't fully constructed from the Killing fields, and that the Killing field here stands for Killing field of the background metric only, not of the full metric (background + perturbation).
The twist \eqref{radialtwist} is expressed in spherical coordinates $(t, r, \vartheta, \varphi)$. In these coordinates, there are two non-vanishing commutators:
\begin{equation} \label{rt-commutator}
	\begin{split}
		[t\stackrel{\star}{,} r] &= i a \alpha A(r), \\
		[\varphi \stackrel{\star}{,} r] &= i a \beta A(r).
	\end{split}
\end{equation}
As shown in~\cite{Schenkel:2011biz}, any Abelian twist, including \eqref{radialtwist}, can be written in Moyal form by performing a coordinate transformation adapted to the twist.
In this new coordinate system, the right-hand side of the commutators becomes constant, unlike in \eqref{rt-commutator}.
The required canonical coordinates are obtained by replacing $r$ with $\hat r$, defined through
\begin{equation} \label{hatrdef}
	\frac{d \hat r}{d r} = \frac{1}{A(r)}.
\end{equation}
We will require this transformation to be well-defined for $r > R$, thus $A(r)$ should be nonzero in this region and possibly zero only at the horizon $r = R$.
The twist and the $\star$-product in terms of $\hat r$ coordinate are given by
\begin{align}
	\mathcal{F} &= \exp \frac{-i a }{2}\Big( (\alpha \partial_t + \beta \partial_\varphi) \otimes \partial_{\hat r} - \partial_{\hat r} \otimes (\alpha \partial_t + \beta \partial_\varphi) \Big), \\
	f \star g &= f\ \exp \frac{i a}{2}\Big((\alpha \overleftarrow{\partial}_t + \beta \overleftarrow{\partial}_\varphi) \overrightarrow{\partial}_{\hat r} - \overleftarrow{\partial}_{\hat r} (\alpha \overrightarrow{\partial}_t + \beta \overrightarrow \partial_\varphi) \Big) \ g. \label{rhatstarprod}
\end{align}
The commutation relations are now of Moyal type (canonical), 
\begin{equation} 
	\begin{split}
		[t\stackrel{\star}{,} \hat r] &= i a \alpha, \\
		[\varphi \stackrel{\star}{,} \hat r] &= i a \beta.
	\end{split}
\end{equation}

From now on, we will work in coordinates $(t, \hat r, \vartheta, \vp)$. It is important to remember that $\hat r$ is always connected to a specific $A(r)$.
In the next section, we will study the perturbations of the Schwarzschild metric
\begin{equation}
d s^2=-\left(1-\frac{R}{r}\right) c^2 d t^2+\frac{1}{1-R / r} d r^2+r^2\left(d \vartheta^2+\sin ^2 \vartheta d \vp^2\right), \qquad
	R \equiv \frac{2 G M}{c^2}.
\end{equation}

When coupling fields to the background metric, or when perturbing the metric itself, we often have to evaluate $\star$-products between some function of the metric (such as its derivative, or determinant) and the field. 
The Schwarzschild metric is static and spherically symmetric, so the functions of it depend only on $r$ and $\vartheta$ and there is no $t$ or $\varphi$ dependence. 
Since the $\star$-product \eqref{rhatstarprod} has no $\vartheta$ dependence, functions of $\vartheta$ are multiplied by the standard pointwise product. Thus, only the $r$-dependent parts of the Schwarzschild background are nontrivially affected by the $\star$-product.

In the context of black hole perturbation theory, we typically deal with $\star$-products between the perturbation mode, which is of generic form $h(r, \vartheta)e^{i m \varphi - i \omega t}$, and the Schwarzschild-derived part of the generic form $f(r, \vartheta)$. After expressing $r$ in terms of $\hat r$, we can evaluate the $\star$-product between such terms using \eqref{rhatstarprod},
\begin{equation} \label{translation}
\begin{split}
	f(\hat r, \vartheta) \star h(\hat r, \vartheta) e^{i m \varphi - i \omega t} &= f(\hat r, \vartheta) \exp \frac{i a}{2}\Big((\alpha \overleftarrow{\partial}_t + \beta \overleftarrow{\partial}_\varphi) \overrightarrow{\partial}_{\hat r} - \overleftarrow{\partial}_{\hat r} (\alpha \overrightarrow{\partial}_t + \beta \overrightarrow \partial_\varphi) \Big) h(\hat r, \vartheta)e^{i m \varphi - i \omega t} \\
	&= f(\hat r, \vartheta) \exp \frac{i a}{2}\Big(- \overleftarrow{\partial}_{\hat r} (\alpha \overrightarrow{\partial}_t + \beta \overrightarrow \partial_\varphi) \Big) h(\hat r, \vartheta)e^{i m \varphi - i \omega t} \\
	&=f(\hat r, \vartheta)\exp (\frac{ i a}{2}\overleftarrow{\partial}_{\hat r} (-i \lambda) )h(\hat r, \vartheta)e^{i m \varphi - i \omega t} \\
	&=f(\hat r + \frac{\lambda a}{2}, \vartheta)h(\hat r, \vartheta)e^{i m \varphi - i \omega t} ,
\end{split}
\end{equation}
where we introduced $\lambda = -\alpha \omega + \beta m$.
Similarly, one can obtain
\begin{equation}
	h(\hat r, \vartheta) e^{i m \varphi - i \omega t} \star f(\hat r, \vartheta) = h(\hat r, \vartheta)e^{i m \varphi - i \omega t} 	f(\hat r - \frac{\lambda a}{2}, \vartheta).
\end{equation}
In conclusion, the $\star$-product between the Schwarzschild-derived function and the perturbation mode reduces to the usual pointwise product with the Schwarzschild-derived function translated by $\pm \lambda a/2$. The sign is $+$ when this function is on the left and $-$ when it is on the right.
This way of evaluating $\star$-products is reminiscent of Bopp shifts in phase-space quantization~\cite{Curtright:2014sli}. 
In the next section, we will apply it to the setting of noncommutative differential geometry to study noncommutative metric perturbations of the Schwarzschild black hole.

\section{Perturbing the Schwarzschild metric} \label{sec3}
The formalism of noncommutative differential geometry based on Drinfeld twists was largely developed in~\cite{Aschieri:2005zs, Aschieri:2005yw}. Summary of this formalism, along with application to black hole perturbation theory up to first order in NC parameter is presented in~\cite{Herceg:2023pmc, Herceg:2024vwc}.
In this section, we study perturbation theory of the Schwarzschild black hole in noncommutative spacetime characterized by the commutation relations \eqref{rt-commutator}.\\

Working with the coordinate $\hat r$ defined in \eqref{hatrdef} simplifies the twist \eqref{radialtwist}, it renders the commutation relations canonical and allows for the $\star$-products to be evaluated as translations.
It is then convenient to use the basis
$(\partial_t, \partial_{\hat r}, \partial_\vartheta, \partial_\varphi)$ for vector fields and the dual basis
$(dt, d\hat r, d\vartheta, d\varphi)$ for 1-forms.
Every element of this basis commutes with the vector fields $\partial_{\hat r}, \partial_t, \partial_\varphi$ that generate the twist, and hence with all functions in the algebra. This makes it a central basis\footnote{The so-called nice basis of~\cite{Schenkel:2011biz}.}, which greatly simplifies computations.
By contrast, using $\partial_r$ instead of $\partial_{\hat r}$ would break centrality since
$[ \partial_r, \partial_{\hat r}] = [\partial_r, A(r)\partial_r] \neq 0$.\\

The Schwarzschild metric in this basis is given by
\begin{equation} \label{schwarzschildnew}
	d s^2=-\left(1-\frac{R}{r}\right) c^2 d t^2+A(r)^2\frac{1}{1-R / r} d \hat r^2+r^2\left(d \vartheta^2+\sin ^2 \vartheta d \vp^2\right).
\end{equation}
The $r$ coordinate can always be expressed in terms of $\hat r$ using \eqref{hatrdef}. We will study the linearized gravitational perturbation theory of this metric by introducing the perturbation $h$ and discarding terms of the order $h^2$ in the calculation. Therefore the full metric is 
\begin{equation}
	g_{\mu \nu} = \oo{g}_{\mu \nu} + h_{\mu \nu},
\end{equation}
where $\oo{g}_{\mu \nu}$ is the Schwarzschild metric \eqref{schwarzschildnew}. Due to the spherical symmetry of the background, the perturbation $h_{\mu \nu}$ can be decomposed into tensor spherical harmonics and seven radial functions as demonstrated by Regge and Wheeler~\cite{Regge:1957td}. 
Furthermore, this decomposition can be grouped into two parts of different parity with respect to the antipodal inversion operator, axial ($-1$) and polar ($+1$). In this paper, we will be interested in the axial sector.

Axial perturbation mode in the Regge-Wheeler gauge is given by
\begin{equation}
\begin{array}{ll}
	h_{t \vartheta}=\frac{1}{\sin \vartheta} \sum\limits_{\ell, m} h_0^{\ell m}(\hat r) \partial_{\varphi} Y_{\ell m}(\vartheta, \varphi)e^{-i \omega t}, & h_{t \varphi}=-\sin \vartheta \sum\limits_{\ell, m} h_0^{\ell m}(\hat r) \partial_\vartheta Y_{\ell m}(\vartheta, \varphi)e^{-i \omega t}, \\
	h_{\hat r \vartheta}= \frac{A(r)}{\sin \vartheta} \sum\limits_{\ell, m} h_1^{\ell m}(\hat r) \partial_{\varphi} Y_{\ell m}(\vartheta, \varphi) e^{-i \omega t}, & h_{\hat r \varphi}=-A(r)\sin \vartheta \sum\limits_{\ell, m} h_1^{\ell m}(\hat r) \partial_\vartheta Y_{\ell m}(\vartheta, \varphi) e^{-i \omega t},
\end{array}
\end{equation}
where we consider a single frequency mode $e^{- i \omega t}$. This single-mode parametrization is justified by the non-existence of a coupling between modes with different $\omega$ or polar/axial coupling at the linear level in $h$. 
The metric $g_{\mu \nu}$ has a $\star$-inverse given by the formula
\begin{equation} 
	g^{\star \mu \nu} = \oo{g}^{\mu \nu} - \oo{g}^{\mu \alpha} \star h_{\alpha \beta} \star \oo{g}^{\beta \nu}.
\end{equation}
We are now in a position to calculate the NC metric inverse, connection coefficients, curvature tensors, and finally the $\mathcal{R}$-symmetrized NC Einstein tensor given by~\cite{Herceg:2023zlk, Herceg:2023pmc}

\begin{equation}  \label{einstein}
\hat{{\rm R}}_{\mu\nu}\equiv\frac{1}{2}\left\langle dx^{\alpha}, \hat{R}(\partial_{\alpha}, \partial_{\mu}, \partial_{\nu})+\hat{R}(\partial_{\alpha}, \bar{R}^{A}(\partial_\nu), \bar{R}_{A}(\partial_\mu))\right\rangle_\star.
\end{equation}
This is an $\mathcal{R}$-symmetrization\footnote{The $\mathcal R$-matrix induced from the twist $\mathcal{F}$ reflects the braiding present in this formalism. In the expression for the NC Einstein tensor, we use the $\mathcal R$-matrix inverse ${\mathcal R}^{-1} = \bar{R}^A \otimes \bar{R}_A.$ and $\star$-pairing of vector fields and 1-forms $\langle \ , \ \rangle_{\star}$.} of the NC Ricci tensor from~\cite{Aschieri:2005zs, Aschieri:2005yw} and can be thought of as a kind of Hermitization. 
The whole calculation, starting from the $\star$-inverse metric up to the NC Einstein tensor must be done using $\hat r$ in place of $r$ by utilizing \eqref{hatrdef}. Therefore, all functions are implicitly assumed to be functions of $\hat r$ and radial derivatives are assumed to be $\partial_{\hat r}$ as well, so we write $h' = \partial_{\hat r} h$.
The formulas for NC metric inverse, Levi-Civita connection, curvature and Einstein tensor simplify in the central basis $\partial_\mu \in \{\partial_t, \partial_{\hat r}, \partial_\vartheta, \partial_\varphi \}$ and $dx^\mu \in \{d t, d {\hat r}, d \vartheta, d \varphi \}$:
\begin{equation}
\begin{split}
g^{\mu \nu}_\star \star g_{\nu \rho} &= \delta^\mu_{~ \rho}, \quad g_{\mu \nu} \star g^{\nu \rho}_\star = \delta_{\mu}^{~ \rho},\\
  \hat{\Gamma}^{\mu}_{~\nu \rho} &= \frac{1}{2} g^{ \mu \alpha}_\star \star \big( \partial_{\nu} g_{\rho \alpha}  +  \partial_{\rho} g_{\nu \alpha} -  \partial_{\alpha} g_{\nu \rho} \big),  \\
   \hat{R}^{~~~~\sigma}_{\mu \nu \rho} &= \partial_{\mu} \hat{\Gamma}^{\sigma}_{~\nu \rho} - \partial_{\nu} \hat{\Gamma}^{\sigma}_{~\mu \rho} + \hat{\Gamma}^{\beta}_{~\nu \rho} \star {\hat \Gamma}^{\sigma}_{~\mu \beta} - {\hat \Gamma}^{\beta}_{~\mu \rho}  \star {\hat \Gamma}^{\sigma}_{~\nu \beta},\\
   \hat R_{\mu \nu} &= \hat{R}^{~~~~\rho}_{\rho \mu \nu},\\
   	\hat{{\rm R}}_{\mu\nu}&=\hat{R}_{(\mu\nu)} \equiv \frac{1}{2} (\hat{R}_{\mu\nu} + \hat{R}_{\nu\mu}).
\end{split}
\end{equation}

When calculating $\star$-products between the Schwarzschild-derived functions and perturbation, the Schwarzschild-derived function undergoes translation as illustrated in \eqref{translation}.
There are two radially dependent functions in the Schwarzschild metric, $r$ and $A(r)$, whose translations in $\hat r$ are elegantly tackled by introducing the notation
\begin{align}
	r_\pm &= r(\hat r \pm \lambda a / 2), \\
	A_\pm &= A(r_\pm).
\end{align}
Despite having 10 components, the vacuum NC Einstein equation $\hat{{\rm R}}_{\mu \nu} = 0$ results in only three distinct radial equations.
The $\hat{{\rm R}}_{t \varphi} = 0$ equation reduces to
\begin{equation} \label{Rtphi}
	B_1 h_1 + B_2 h_1' + B_3 h_0 + B_4 h_0' + B_5 h_0'' = 0,
\end{equation}
with coefficients $B_i$ given in the equation \eqref{RtphiC} of Appendix \ref{appendix}.
The $\hat{{\rm R}}_{r \varphi} = 0$ equation is
\begin{equation} \label{Rrphi}
	C_1 h_1 + C_2 h_1' +C_3 h_0 +C_4 h_0' = 0,
\end{equation}
with coefficients  $C_i$ given in the equation \eqref{RrphiC} of Appendix \ref{appendix}.
The $\hat{{\rm R}}_{\vartheta \varphi} = 0$ equation is
\begin{equation}
	\begin{split} \label{Rthphi}
        \Biggl( -\frac{r_+ \left(R - r_+\right) \Bigl( A_- \left(r_- A' + 2 A r_-'\right) 
        + A r_- A_-'\Bigr)}{A_- A_+^2 r_-} + \frac{A \left(R - r_-\right) r_-'}{A_-^2} \\
        + \frac{2 A r_+ A_+' \left(R - r_+\right)}{A_+^3} + \frac{A \left(2 R - r_+\right) r_+'}{A_+^2} \Biggr) h_1
        + \frac{A r_+ \left(r_+ - R\right)}{A_+^2}h_1' - \frac{i r_+^3 \omega}{R - r_+}h_0 = 0.
    \end{split}
\end{equation}
Combining the first-order equations \eqref{Rrphi} and \eqref{Rthphi}, we get a single second-order differential equation in $h_1$, 
\begin{equation} \label{radialeq}
	D_1 h_1 + D_2 h_1' + D_3 h_1'' = 0,
\end{equation}
with coefficients $D_i$ given in equations \eqref{radialC1} and \eqref{radialC2} of Appendix \ref{appendix}.
This equation can be reduced to Schr\"odinger form by introducing the tortoise variable $r_*$ and redefining the field,
\begin{equation}
	\frac{d r_*}{d \hat r} = \xi(\hat r), \qquad h_1(\hat r) = W(\hat r)\psi(\hat r) ,
\end{equation}
where
\begin{align}
	\xi &= \frac{A_+ r_+}{r_+-R}, \label{xi} \\ 
	\frac{W'}{W} &= -\frac{A'}{A}-\frac{A_-'}{2 A_-}+\frac{3 A_+'}{2 A_+}+\frac{\left(2 R-r_+\right) r_+'}{r_+ \left(R-r_+\right)}-\frac{r_-'}{2 r_-}.
\end{align}
The equation \eqref{radialeq} reduces to the Schr\"odinger-like form
\begin{equation} \label{master1}
	\frac{d^2 \psi}{d r_*^2} + \big( \omega^2 - V \big) \psi = 0,
\end{equation}
where the effective potential is given by
\begin{equation}
	\begin{split} \label{master2}
	V = \frac{1}{4 r_+^4} \Bigg(
& -\frac{(R - r_+) (4 A_+^4 \ell (\ell + 1) r_-^2 r_+ (R - r_-) + A_+^2 r_+^3 ((17 R - 15 r_-) (r_-')^2 + 10 r_- (r_- - R) r_-'')}{A_+^4 r_-^2 (R - r_-)} \\
& + \frac{A_+^2 \Big(8 r_-^2 R (r_- - R) (r_+')^2 + 2 r_- r_+ r_+' (R (5 R - 4 r_-) r_-' + 2 r_- (R - r_-) r_+') \Big)}{A_+^4 r_-^2 (R - r_-)} \\
& + \frac{2 A_+ r_- r_+ (R - r_-) \Big(A_+' (2 r_+ (r_+ - R) r_-' + r_- (4 r_+ - 3 R) r_+') + r_- r_+ A_+'' (r_+ - R)\Big)}{A_+^4 r_-^2 (R - r_-)} \\
& + \frac{3 r_-^2 r_+^2 (A_+')^2 (R - r_-) (R - r_+)}{A_+^4 r_-^2 (R - r_-)} \\
& + \frac{2 r_+ (R - r_+) (A_-' (3 r_+ (R - r_+) r_-' + r_- (4 r_+ - 3 R) r_+') + r_- r_+ A_-'' (r_+ - R))}{A_- A_+^2 r_-} \\
& + \frac{(R - r_+) (r_- (2 A_+ r_+ (A_+ (8 (r_-')^2 - R r_-'') + 2 R A_+' r_-') + 3 r_+^2 R (A_-')^2 + 12 A_+^2 R r_-' r_+' - 3 r_+^3 (A_-')^2))}{A_-^2 A_+^2 r_-} \\
& - \frac{16 A_+^2 r_+ R (r_-')^2 + 2 A_+ r_-^2 (A_+ r_+ r_-'' - 2 r_-' (r_+ A_+' + 3 A_+ r_+'))}{A_-^2 A_+^2 r_-} \\
& + \frac{A_+^2 (R - r_-)^2 (r_-')^2}{A_-^4} - \frac{2 r_+ A_-' (R - r_-) (R - r_+) r_-'}{A_-^3}
\Bigg).
	\end{split}
\end{equation}
This is the master formula that gives the analytic expression for the effective potential up to all orders in noncommutativity parameter for an arbitrary $A(r)$.
In the limit $a \to 0 \implies A = 1, \hat r = r$, the Regge-Wheeler potential is restored,
\begin{equation*}
	\lim_{a \to 0} V = \frac{(r-R) (\ell(\ell + 1) r-3 R)}{r^4}.
\end{equation*}

\section{Some examples} \label{sec4}

\subsection{The Moyal space in $\mathbf{r - (t, \varphi );\ A = 1}$}
\begin{figure}[!h]
	\subfigure[]{\includegraphics[scale=0.40]{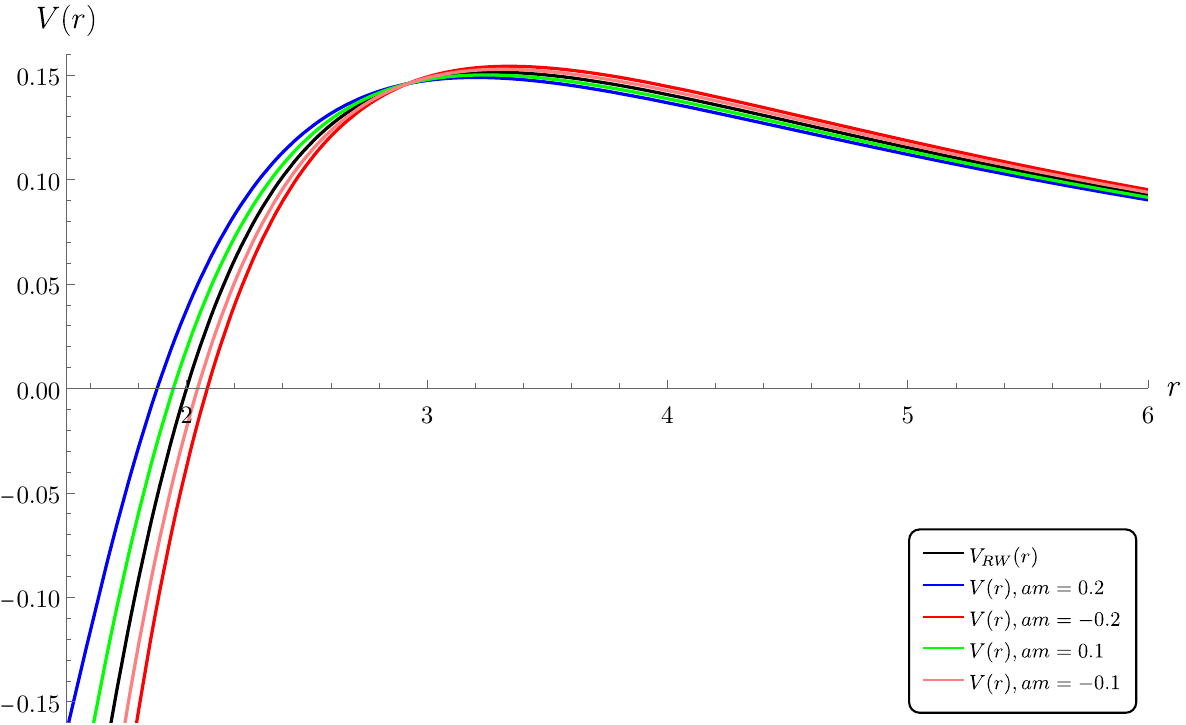}\label{first_order}} \qquad
	\subfigure[]{\includegraphics[scale=0.40]{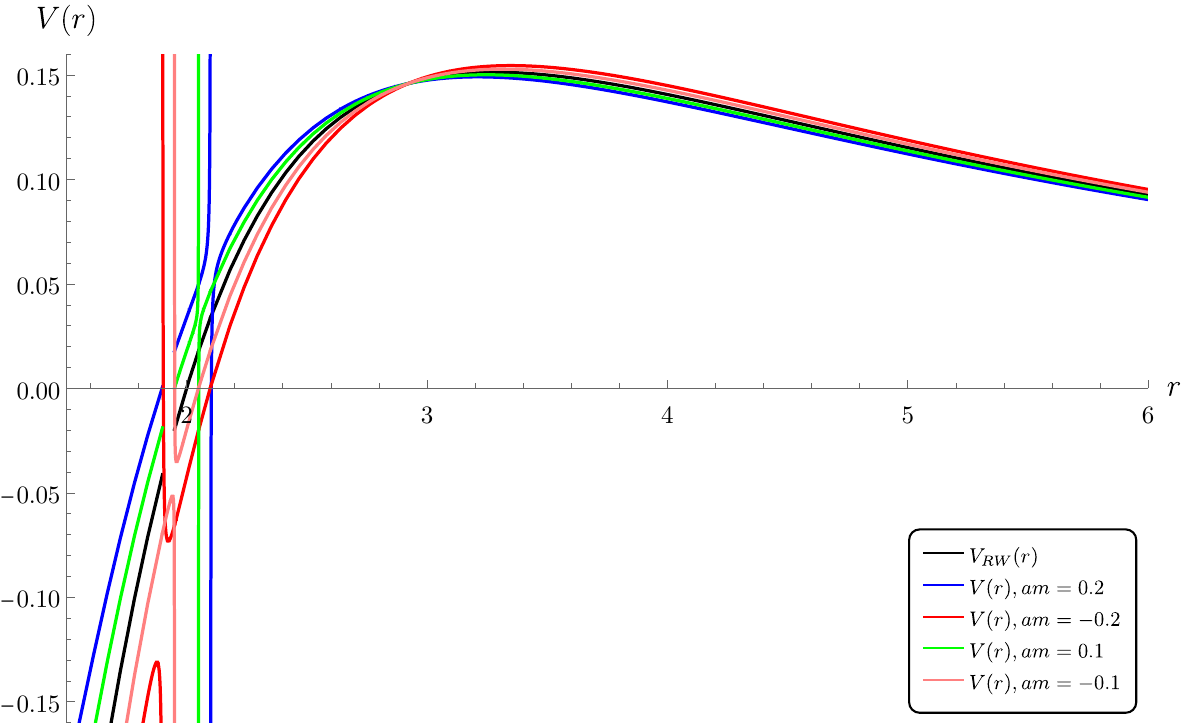}\label{nonpert1}}  
	\caption{Noncommutative potentials for the $\ell = 2, M = 1$ case and $A = 1$. First-order in $a$, given in \eqref{pot0}, is on the left and up to all orders, given in \eqref{pot1}, is on the right.} \label{fig1}
\end{figure}

By choosing $A = 1$ and expanding only up to the first order in $a$, we obtain the same potential as in~\cite{Herceg:2023pmc},
\begin{equation} \label{pot0}
	V = \frac{(r-R) (\ell(\ell + 1) r-3 R)}{r^4} + \lambda a \frac{  \ell(\ell + 1) r (3 R-2 r)+R (5 r-8 R)}{2 r^5}.
\end{equation}
A generalization of this potential to all orders in $a$ is given by

\begin{equation}
	\begin{split} \label{pot1}
		V = \frac{1}{4 r_-^2 r_+^4 (R - r_-)} \Bigg[
    & r_-^3 \Big(4 (\ell(\ell + 1) + 5) r_+ R - 4 (\ell(\ell + 1) - 3) r_+^2 - 35 R^2\Big) + r_-^2 \Big(4 r_+ \Big((\ell(\ell + 1) - 13) r_+ R \\ 
	& - (\ell(\ell + 1) - 4) R^2 + 4 r_+^2\Big)  + 21 R^3\Big)  - r_+ r_- (R - r_+) \Big(26 R^2 - 3 r_+ (5 r_+ + R)\Big) \\
		& + 3 r_-^4 (5 R - 4 r_+) + 17 r_+^2 R (R - r_+)^2  - r_-^5 \Bigg],
\end{split}
\end{equation}
where $r_\pm = r \pm \lambda a / 2$.
This potential is plotted in Fig. \ref{fig1} for the case $\alpha = 0, \beta = 1 \implies \lambda = m$. The poles are located at $r_- = R,\ r_- = 0$ and $r_+ = 0$. This is in contrast with the classical case, where only a single pole appears at $r = 0$.

The near-horizon behavior of the potential \eqref{pot0} for the choice of parameters $\alpha = 0, \beta = 1$ was analyzed in~\cite{Herceg:2024upt}. It has been shown that the translated radial coordinate $r_+ = r + a m / 2$ displays no pathological behavior when relating it to the tortoise coordinate, since the Jacobian is exactly the same as the commutative one with $r_+$ instead of $r$:
\begin{equation}
	\frac{d r_*}{d r_+} = \frac{r_+}{r_+ - R}.
\end{equation}
In~\cite{Herceg:2024upt}, the observation was made by considering only the first order in $a$. Now we can confirm that this relation is exact to all orders as instructed by the formula \eqref{xi} when substituting $A = 1,\ \hat r = r \implies r_+ = r + a m / 2, \ \partial_r = \partial_{r_+}$. This also leads to higher-order corrections to the asymptotic value of the potential at the horizon (which is located at $r_+ = R$):
\begin{equation}
	V(r_+ = R) = \frac{a^2}{4 R^4}.
\end{equation}

From Fig. \ref{nonpert1}, we can see the potential's divergences that appear close to the horizon. These are located at $r = R - a m / 2$. 
These divergences appear starting at the third order in the noncommutativity parameter $a$ as can be seen from the Taylor expansion of the potential \eqref{pot1}:
\begin{equation}
	\begin{split}
		V &= \frac{(r-R)(\ell(\ell + 1)r - 3 R)}{r^4} + \lambda a \frac{(5r - 8R)R + \ell(\ell + 1) r (3R - 2r)}{2r^5} \\
		&+ (\lambda a)^2\frac{3 (\ell(\ell + 1)+10) r^2-2 (3
		\ell(\ell + 1)+29) r R+35 R^2}{4 r^6} \\
		&+ (\lambda a)^3 \frac{\ell(\ell + 1) \left(-2 r^3+7 r^2 R-5 r R^2\right)-16
   r^3+59 r^2 R-78 r R^2+33 R^3}{4 r^7 (r-R)}.
	\end{split}
\end{equation}
Divergences arise from $\star$-products of functions that are classically divergent at the horizon. To clarify, we will consider a $\star$-product 
\begin{equation*}
	\frac{1}{1-\frac{R}{r}} \star e^{i m \varphi} = \frac{1}{1 - \frac{R}{r +  a m / 2}}e^{i m \varphi}.
\end{equation*}
The function $\frac{1}{1-R/r}$ is divergent at the point $r = R$, but its translated counterpart is divergent at a different point $r = R - a m / 2$. The Regge-Wheeler potential is classically zero at the horizon, which implies that all such functions are multiplied by their multiplicative inverse at some point, thereby removing the divergence.
In the NC case, the supposed inverse might arise from an oppositely translated $\star$-product. Thus, if in the commutative case we have
\begin{equation*}
	\frac{1}{1 - R/r} \big(1 - R/r\big) = 1,
\end{equation*}
then in the NC case we might have
\begin{equation} \label{argument}
\frac{1}{1 - \frac{R}{r + a m / 2}}\Big(1 - \frac{R}{r - a m / 2}\Big) = \frac{(a m / 2+ r) (a m / 2- r+ R)}{(a m / 2- r) (a m / 2+ r- R)},
\end{equation}
which is clearly divergent at $r = R - a m / 2$ and $r = a m / 2$.
In the case of $r-t$ noncommutativity, where $\alpha = 1, \beta = 0 \implies \lambda = -\omega$, $m$ in the above expression is replaced by $- \omega$.
On-shell frequency $\omega$ is complex when we solve for quasinormal mode (QNM) boundary conditions; therefore the imaginary component of frequency will soften this divergence into finite Lorentzian-like peaks with width proportional to the magnitude of the imaginary part. In other words, the poles move away from the real line into the complex plane.\\

There is, however, a class of $A(r)$ that leaves the horizon invariant and does not exhibit any non-classical divergence (even for real $\omega$). This seems to be the case when $A(r)$ smoothly vanishes at the horizon. One example is the choice $A(r) = 1-R/r$.

\subsection{The Moyal space in $\mathbf{r_* -  t;\ A = 1 - R/r}$}
\begin{figure}[!h]
\includegraphics[scale=0.60]{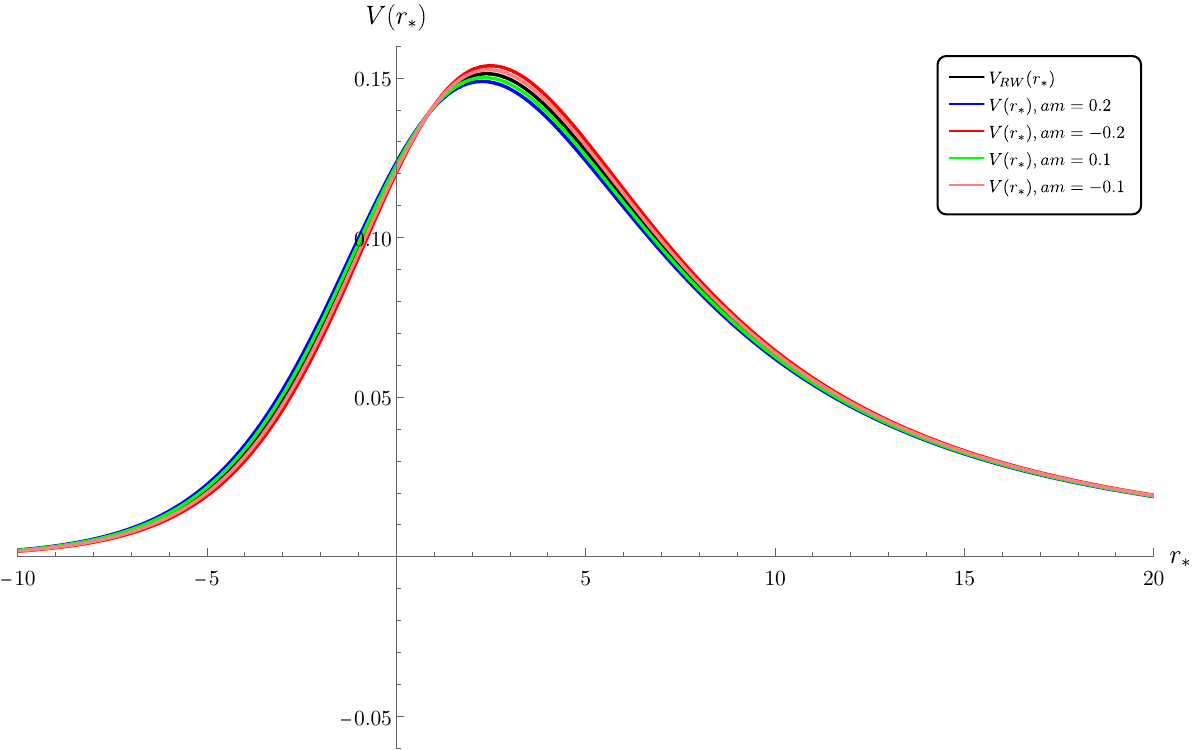}
	\caption{Noncommutative potentials with $A(r) = 1-R/r$ for the $\ell = 2, M = 1$ case.} \label{fig2}
\end{figure}

The choice
\begin{equation}
	A(r) = 1-R/r
\end{equation}
leads to
\begin{equation}
	\hat r = r_* = r + R \log (r/R - 1),
\end{equation}
the usual Schwarzschild tortoise coordinate. It is also equivalent to the tortoise coordinate as defined in \eqref{xi} since $\xi = 1$. The inverse relation is
\begin{equation}
	r(r_*) = R + R\ \mathcal{W}(e^{r_*/R-1}),
\end{equation}
where $\mathcal{W}$ is the Lambert $\mathcal{W}$ function. This choice of $A(r)$ leads to noncommutativity of the form
\begin{equation}
	[t \stackrel{\star}{,} r_*] = i a.
\end{equation}
The potential is
\begin{equation}
	\begin{split}
		V &= \frac{1}{4 r_-^4 r_+^6} \Bigg[
	r_+^2 r_-^4 \Big(-4 (\ell(\ell + 1)- 2) r_+ R + 4 (\ell(\ell + 1) - 3) r_+^2 + 5 R^2\Big)
     - 4 r_+^3 r_-^3 \Big(-6 r_+ R + 4 r_+^2 + 3 R^2\Big)  \\
    & + r_+^4 r_-^2 \Big(5 r_+ \Big(3 r_+ + 2 R\Big) - 8 R^2\Big) 
     + 18 r_+^6 R^2 
     + r_-^6 \Big(R - r_+\Big)^2 
     + 4 r_+ r_-^5 \Big(2 R - 3 r_+\Big) \Big(R - r_+\Big) 
     - 32 r_+^6 r_- R
		\Bigg],
	\end{split}
\end{equation}
where $r_\pm = r(r_* \pm \lambda a / 2)$. The plot of the potential for the case $\alpha = 0, \beta = 1$ is given in Fig. \ref{fig2}.

\clearpage

\subsection{$\mathbf{\kappa}$-Minkowski space in $\mathbf{r-t; \ A = r/R}$}
\begin{figure}[H]
	\subfigure[]{\includegraphics[scale=0.55]{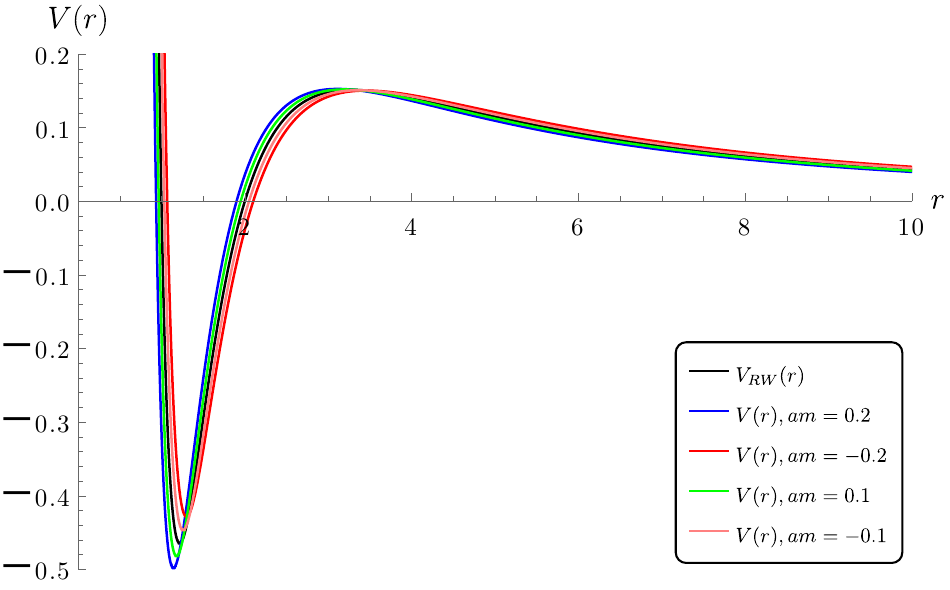}\label{kappa}} \qquad
		\subfigure[]{\includegraphics[scale=0.55]{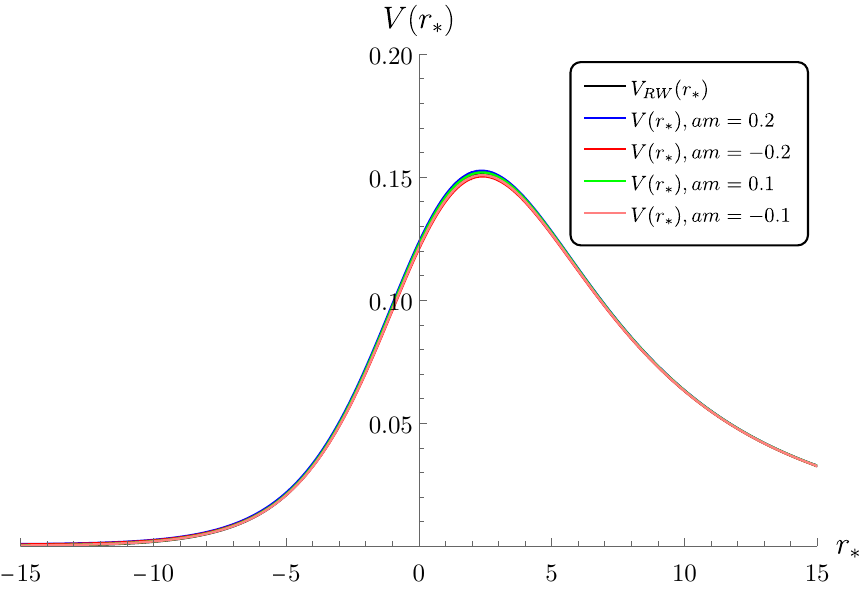}\label{kappatort}}  \\
	\caption{Potentials for $A(r) = r/R, \alpha = 0, \beta = 1.$}
\end{figure}

Another interesting case is $A(r)=r/R$ with $\alpha = 1, \beta = 0$. This leads to noncommutativity of the form
\begin{equation}
		[t \stackrel{\star}{,} r] = i a r/R,
\end{equation}
known as the $\kappa$-Minkowski spacetime~\cite{Majid:1994cy}. Effective potential in this case is
\begin{equation}
	V = \frac{e^{-\frac{2 a \omega }{R}} \left(-4 e^{\frac{a \omega }{R}} \left(R^2-\ell(\ell + 1) r^2\right)-2 (2 \ell(\ell + 1)+7) r R e^{\frac{a \omega }{2 R}}+2 r
   e^{\frac{3 a \omega }{2 R}}+R^2 e^{\frac{2 a \omega }{R}}+15 R^2\right)}{4 r^4}.
\end{equation}
There are no divergences at the horizon in this case either, even though applying the same argument as in \eqref{argument} suggests they are expected. 
The exponential terms in the potential are a consequence of  the dilating nature of $\kappa$-Minkowski spacetime.
The angular version is given by the choice $\alpha = 0, \beta = 1$. The potential is
\begin{equation} \label{angular}
	V = \frac{e^{\frac{2 a m}{R}} \left(-4 e^{-\frac{a m}{R}} \left(R^2-\ell(\ell + 1) r^2\right)-2 (2 \ell(\ell + 1)+7) r R e^{\frac{-a m}{2 R}}+2 r
   e^{\frac{-3 a m}{2 R}}+R^2 e^{\frac{-2 a m}{R}}+15 R^2\right)}{4 r^4}.
\end{equation}
This potential, for various values of $a m$, is plotted in terms of $r$ in Fig. \ref{kappa}, and its $r_*$ coordinate given in \eqref{xi} in Fig. \ref{kappatort}.

\begin{figure}[H]
	\subfigure[]{\includegraphics[scale=0.4]{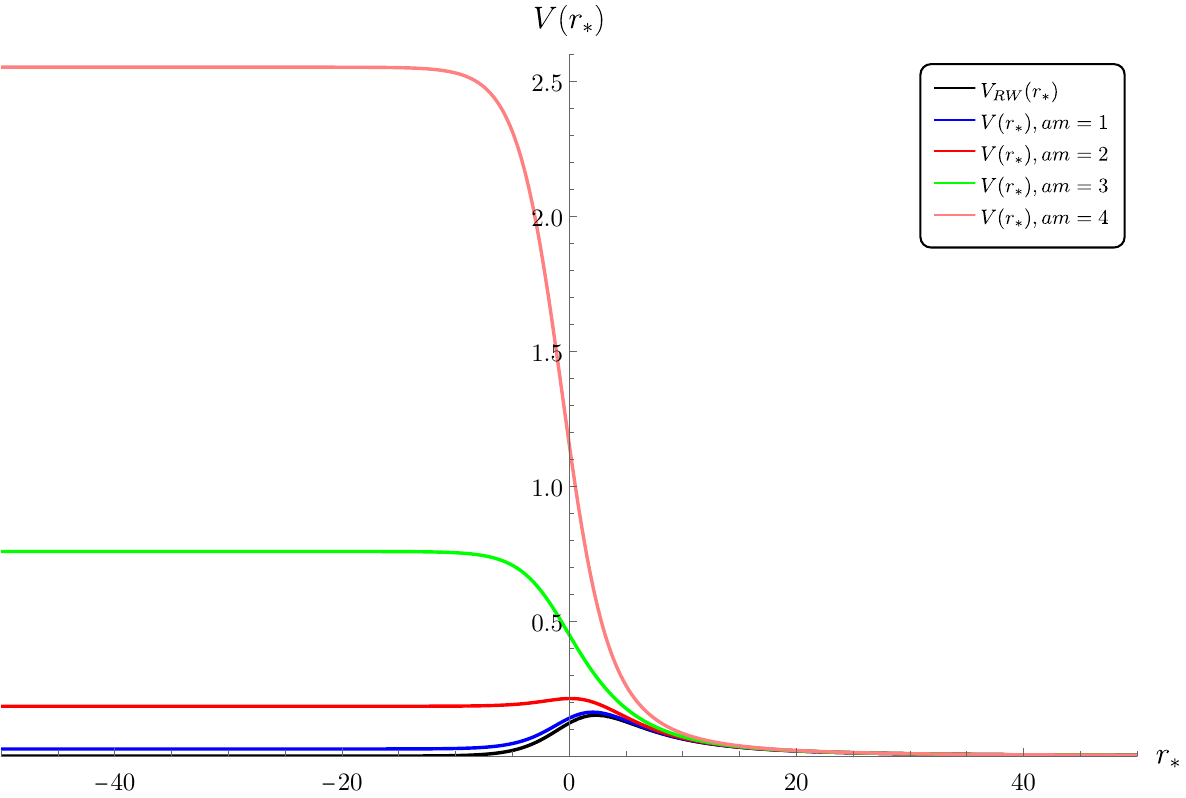}}\quad
	\subfigure[]{\includegraphics[scale=0.4]{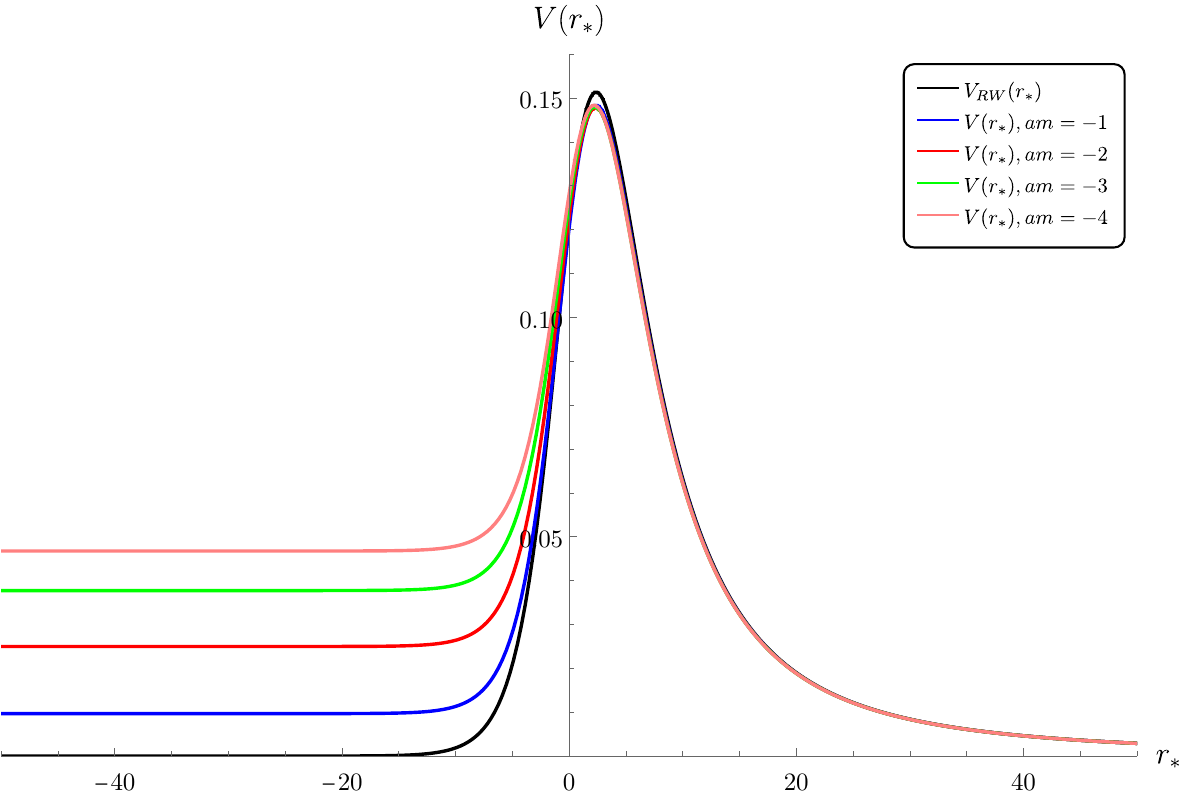}}  \\
	\caption{Potentials for large values of $am$. Positive values are on the left figure and negative on the right.} 
	\label{planck1}
\end{figure}

\begin{figure}[H]
	\subfigure[]{\includegraphics[scale=0.4]{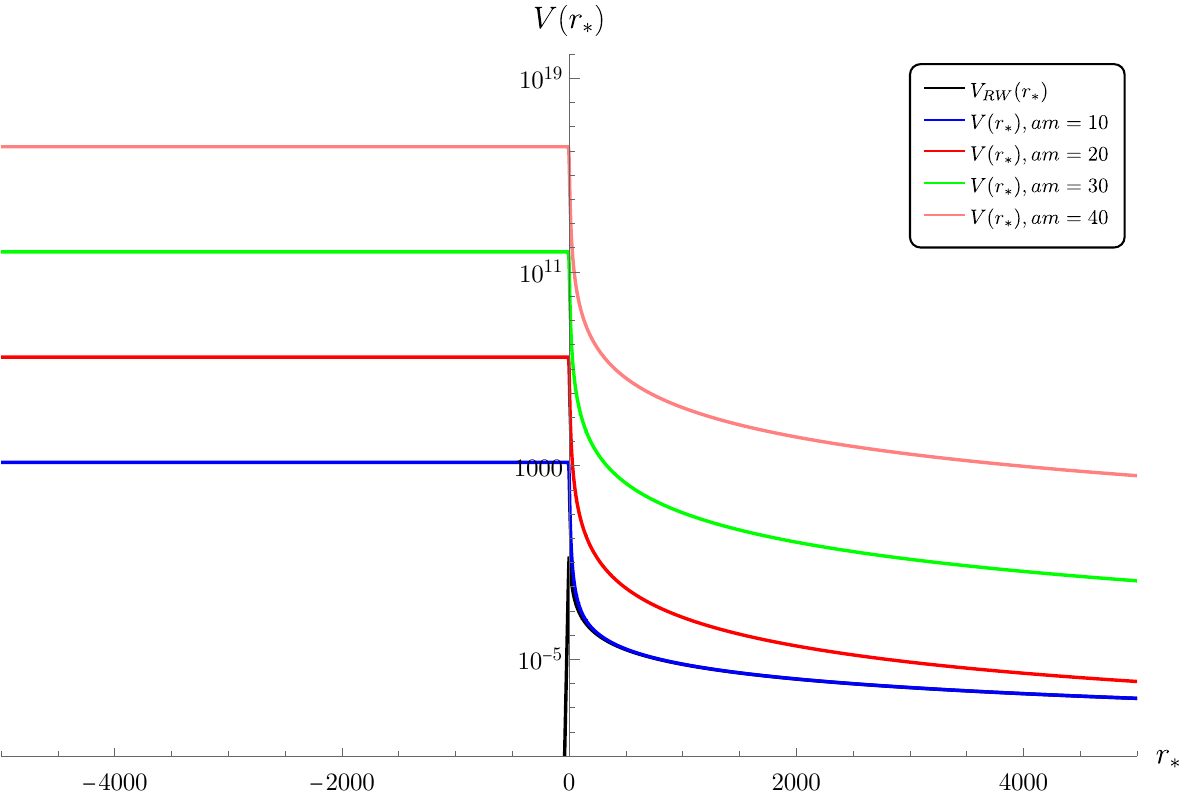}} \quad
	\subfigure[]{\includegraphics[scale=0.4]{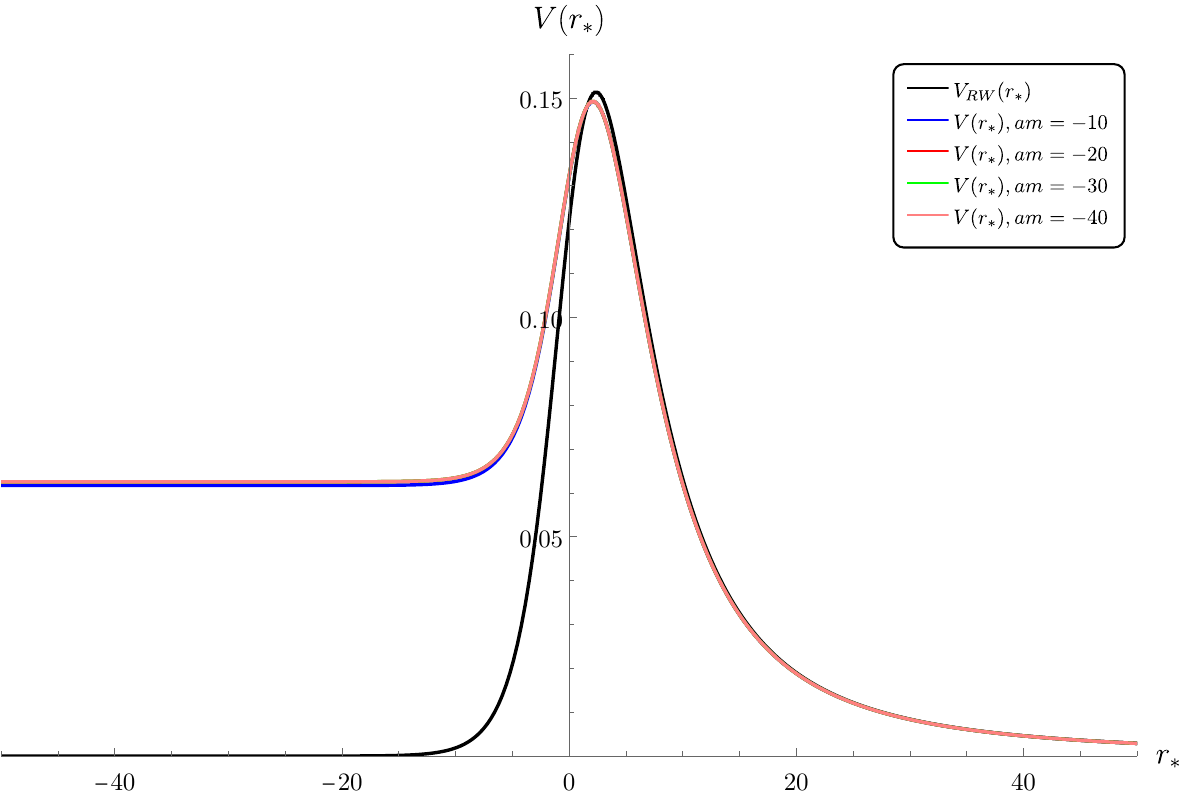}}  \\
	\caption{Potentials $|am| < 50$. Positive values are on the left figure and negative on the right.} 
	\label{planck2}
\end{figure}

Nonperturbative evaluation of $\star$-products enables the study of the regime where the noncommutativity constant is on the same scale as the radius of the black hole. This scenario is relevant for Planck-scale black holes, which are phenomenologically interesting as dark matter candidates.
Potentials in this regime significantly deviate from the single-peak structure typical of macroscopic black holes. 
This is illustrated for the angular version of $\kappa$-Minkowski, i.e. $A(r) = r/R, \alpha = 0, \beta = 1$.
In Fig. \ref{planck1}, it can be seen how large values of $am$ influence the potential.
Both regimes result in positive asymptotic values of the potential to the left of the peak. This value, however, grows exponentially in the case of positive $am$. This behavior is expected given that there are exponential terms in the potential formula \eqref{angular}.
This trend continues for even larger values of $a m$ as can be seen from Fig. \ref{planck2}.
Interestingly, the potential does not change significantly for negative values of $a m$. For positive values, the barrier continues to increase, implying a stronger reflection of the gravitational waves.

\begin{figure}[H]
	\subfigure[]{\includegraphics[scale=0.65]{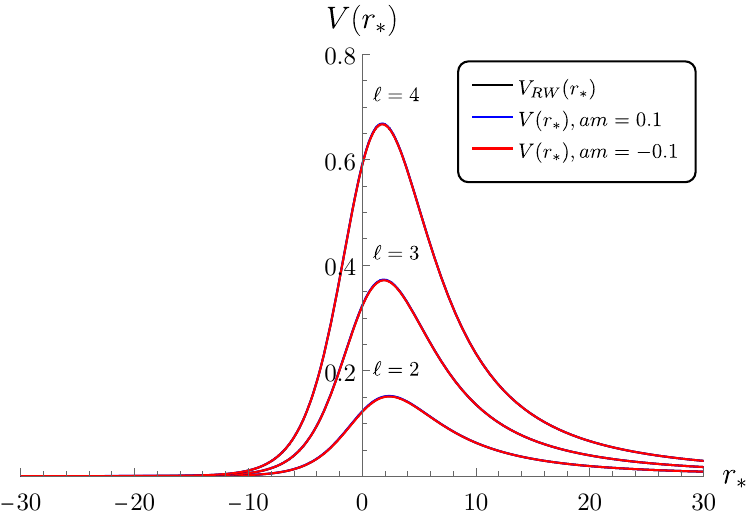}} \quad
	\subfigure[]{\includegraphics[scale=0.65]{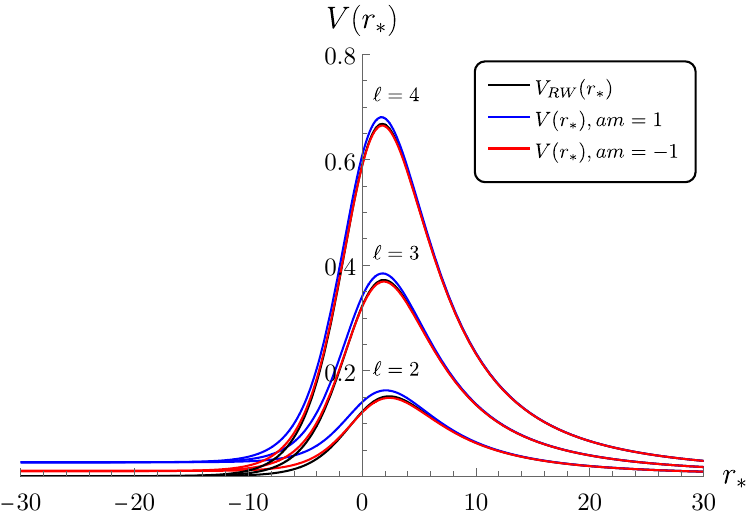}}  \\
	\caption{Potentials of higher $\ell$ modes for $A = r/R$.}
	\label{figell}
\end{figure}

In Fig. \ref{figell}, we can see the behavior of the potential for higher $\ell$ modes.

\section{Conclusion} \label{conclusion}

The potency of nonperturbative results obtained here may be best seen in situations where quantum gravity effects are so strong  that all conclusions gained from just the first few orders in perturbative calculations become unreliable. 
In extreme situations like these, where the perturbation analysis fails and is not to be trusted anymore, the approach which is exact and nonperturbative in the deformation parameter apparently  establishes the only appropriate way out for inferring  relevant  information regarding the black hole relaxation dynamics.\\

These extreme situations, for example, include strong-gravity regimes and objects like primordial  black holes where  quantum corrections  are so large as to break the limits within which the perturbative treatment  is considered reliable.
Unlike the supermassive black holes found at the center of galaxies, these black holes  are incredibly small and short-lived\footnote{Under the assumption that no new physics beyond the Standard Model or modifications of gravity are assumed. If some of the mentioned features are included, the situation may dramatically change~\cite{Baker:2025zxm}.}. They are of Planck-scale size and are considered to have been created in the early universe, shortly after the Big Bang, by arising from quantum fluctuations of spacetime, similar to how virtual electron-positron pairs appear and disappear in quantum electrodynamics. These black holes can have very small masses, but due to some limits  associated with the effects of their evaporations  via Hawking  radiation  on big bang nucleosynthesis and the extragalactic photon background, their masses can be estimated to $M \sim 10^9-10^{17} $g~\cite{Carr:2009jm, Young:2014ana}. However, the observational constraints on PBHs across the full mass range are still a field of active research and there is still an asteroid-mass window spanning from $10^{-17}M_{\odot}-10^{-10}M_{\odot}$ which remains unconstrained~\cite{Carr:2026hot}. \\

 From a purely theoretical perspective, Planck-scale black holes are interesting in their own right as they provide a window into the  realm where gravity and quantum mechanics collide, with perhaps the most notable example being the notion of  a fuzzball, a hypothetical object arising from superstring theory, proposed to give a fully quantum description of black holes and resolving two of their major issues, the problem of gravitational singularity  and the black hole information paradox~\cite{Mathur:2005zp, Mathur:2008wi}. Although no experimental evidence for fuzzball conjecture is yet available, their existence might in principle be tested through gravitational-wave astronomy~\cite{Berkowitz:2021bfi}. \\
 
The nonperturbative and noncommutative generalization obtained in this paper for the Regge-Wheeler potential, which includes all orders in the noncommutativity parameter, makes the analysis of  perturbation and stability of black holes, particularly Planck-sized black holes possible. Namely, as already indicated from the perturbative analysis in~\cite{Herceg:2023zlk}, the  BH metrics are stable under NC metric perturbations, consequently affecting the lifetime of these BHs, which is very important for possible detection in the near future. \\

Another extreme situation of interest is the vicinity of a black hole horizon, where it is reasonable to expect significant departures from classical description provided by GR. Indeed, almost all proposed 
 resolutions to problems in quantum gravity, such as 
 the black hole information paradox, rely heavily on  these departures and assume modifications in the near-horizon structures.
It was recently suggested that such modifications to the  near-horizon structures may give rise to late-time echoes in the black hole merger gravitational wave signals obtained by the advanced Laser Interferometer Gravitational-Wave Observatory (LIGO). These gravitational wave signals would  otherwise be indistinguishable from those predicted by GR.\\

Observation of late-time echoes  in the gravitational wave data released by some of the gravitational wave interferometers, and particularly observation of repeating damped echoes with certain time delays, would undoubtedly  establish  the  veracity of the  near-horizon Planck-scale departures from GR inferred from the observed black hole merger events~\cite{Abedi:2016hgu, Agullo:2020hxe}.\\

If not yet observable with the sensitivity provided by the current gravitational wave interferometers, phenomena like late-time echoes might be verified
in observations from future interferometer detectors, which are expected to reach sufficiently high sensitivity. As already pointed out, they could also confirm or rule out  alternatives to classical black holes, such as the fuzzball or firewall paradigms~\cite{Berkowitz:2021bfi, Almheiri:2012rt}.\\

In summary, the main result of this paper is the effective potential related to NC gravitational perturbations in the exact form, including all orders in noncommutative parameter $a$. As the acquaintance with such a result has a decisive role in predicting any of the physical outcomes for the black holes at the Planck scale, our work  may serve as a starting point in this direction. \\

\noindent{\bf Acknowledgment}\\
  This  research was supported by the Croatian Science
Foundation Projects  IP-2020-02-9614 {\it{Search for Quantum spacetime in Black Hole QNM spectrum and Gamma Ray Bursts}}  and IP-2025-02-8625
{\it{Quantum aspects of gravity}}.

\newpage
\appendix

\section{Coefficients of radial equations of motion} \label{appendix}

Here we provide the coefficients of the radial equations of motion \eqref{Rtphi}, \eqref{Rrphi}, and \eqref{radialeq}. Coefficients of the equation \eqref{Rtphi} are

\begin{equation}
	\begin{split} \label{RtphiC}
	B_1 =&\  
        i r_- \omega  \Bigl( A_-^2 r_-^2 \left(R-r_+\right)
        \Bigl( r_+ \left(R-r_+\right) \bigl( A_+ \left(4 r_- A' \left(r_--R\right) + A \left(10 r_--11 R\right) r_-'\right) \\
		&+ 8 A r_- A_+' \left(R-r_-\right) \bigr) + 2 A A_+ r_- R \left(R-r_-\right) r_+' \Bigr) \\
		&- 4 A A_- A_+ r_+ r_-^3 A_-'
        \left(R-r_-\right) \left(R-r_+\right)^2 
	    + A A_+^3 \left(R-r_-\right)^2 \Bigl( 2 r_-^3 \left(R-r_+\right) - r_+^3 R \Bigr) r_-'\Bigr), \\[0.5em]
	B_2 =& - 4 i A A_-^2 A_+ r_+ r_-^4 \omega  \left(R-r_-\right) \left(R-r_+\right)^2,  \\[0.5em]
	B_3 =&\  
        2 A_+ \Bigl( A_-^2 r_-^3 \left(R-r_+\right) 
	    \Bigl( 2 A_+^2 \ell(\ell+1) r_- \left(r_--R\right) - R \left(r_-+r_+-2 R\right) r_-' r_+' \Bigr) \\
	&+ A_+^2 R \left(R-r_-\right) \Bigl( r_-^3 \left(R-r_+\right) + r_+^3 R - r_+^3 r_- \Bigr)
	    \left(r_-'\right)^2 \Bigr), \\[0.5em]
	B_4 =& 
        - \Bigl( 4 A_- A_+ r_+ r_-^4 A_-' \left(R-r_-\right) \left(R-r_+\right)^2 \\
	&+ A_-^2 r_-^3 \left(R-r_+\right) \Bigl( r_+ \left(R-r_+\right)
        \Bigl( A_+ \left(7 R-6 r_-\right) r_-'+8 r_- A_+' \left(r_--R\right) \Bigr) \\
	&- 2 A_+ r_- \left(R-r_-\right) \left(3 R-2 r_+\right) r_+' \Bigr)
	    - A_+^3 r_- \left(R-r_-\right)^2 \Bigl( 2 r_-^3 \left(R-r_+\right)-r_+^3 R \Bigr) r_-'\Bigr), \\[0.5em]
	B_5 =& 
        - 4 A_-^2 A_+ r_+ r_-^4 \left(R-r_-\right) \left(R-r_+\right)^2  .
	\end{split}
\end{equation}

Coefficients of the equation \eqref{Rrphi} are
\begin{equation}
	\begin{split} \label{RrphiC}
	C_1 =&\  
	\frac{2 A' \left(R-r_+\right) \left(r_- r_+'-r_+ r_-'\right)}{A_+^2} 
        + A r_- \Biggl( \frac{\left(4 r_--3 R\right) \left(r_-'\right)^2 + 2 r_- \left(r_--R\right) r_-''}{A_-^2 r_-} \\
	&+ \frac{2 A_-' \left(R-r_-\right) r_-'}{A_-^3} - \frac{2 R A_-' r_+'}{A_- A_+^2} 
	    - \frac{R r_-' r_+'}{A_+^2 \left(R-r_-\right)} - 2 \ell(\ell+1) \Biggr) + \frac{2 A r_- r_+^3 \omega^2}{r_+ - R} \\
	&+ \frac{A r_+ \Bigl( R \left(4 r_- - 3 R\right) \left(r_-'\right)^2 + r_- \left(3 R - 2 r_-\right) r_+' r_-' 
        + 2 r_- R \left(r_- - R\right) r_-'' \Bigr)}{A_+^2 r_- \left(R - r_-\right)} \\
	&+ \frac{A r_+^2 r_-' \biggl( \frac{A_+ \left(3 R - 4 r_-\right) r_-'}{R - r_-} + \frac{2 A_+ r_- A_-'}{A_-} 
        - 6 r_- A_+' \biggr)}{A_+^3 r_-} + \frac{6 A r_+ R A_+' r_-'}{A_+^3} \\
	&+ \frac{2 A r_+ A_-' \left(r_- r_+' - R r_-'\right)}{A_- A_+^2} 
	+ \frac{6 A r_- \left(R - r_-\right) r_-' r_+'}{A_-^2 r_+} + \frac{2 A r_+^2 r_-''}{A_+^2}, \\[0.5em]
	C_2 =&
	- \frac{2 A \left(R - r_+\right) \left(r_+ r_-' - r_- r_+'\right)}{A_+^2}, \\[0.5em]
	C_3 =& - \frac{4 i r_+^3 \omega r_-'}{R - r_+}, \\[0.5em]
	C_4 =&\ \frac{2 i r_- r_+^3 \omega}{R - r_+}.
\end{split}
\end{equation}

Coefficients of the equation \eqref{radialeq} are
\begin{equation}
	\begin{split} \label{radialC1}
		D_1 =& 
 -\frac{A \left( \left(8 r_- - 9 R\right) \left(r_-'\right)^2 + 6 \left(R - r_-\right) r_- r_-'' \right) r_+^4}{\left(R - r_-\right) r_-^2 A_+^2 \left(R - r_+\right)} \\
        &- \frac{A \Bigl( 2 R \left(9 R - 8 r_-\right) \left(r_-'\right)^2 + \left(7 R - 6 r_-\right) r_- r_+' r_-' + 2 r_- \left(r_- - R\right) \left(6 R r_-'' + r_- r_+'' \right) \Bigr) r_+^3}{\left(R - r_-\right) r_-^2 A_+^2 \left(R - r_+\right)} \\
        &+ \frac{12 A \left(r_+ - R\right) \left(A_+'\right)^2 r_+^2}{A_+^4} 
        + \frac{2 \left(r_+ - R\right) A' A_-' r_+^2}{A_- A_+^2} 
        + \frac{2 \left(R - r_+\right) A' \left(4 r_- A_+' - A_+ r_-'\right) r_+^2}{r_- A_+^3} \\
        &+ \frac{2 A \left(r_+ - R\right) A_-'' r_+^2}{A_- A_+^2} 
        + \frac{4 A \left(R - r_+\right) A_+'' r_+^2}{A_+^3} \\
        &- \frac{A \Bigl( R^2 \left(8 r_- - 9 R\right) \left(r_-'\right)^2 + R r_- \left(5 r_- - 7 R\right) r_+' r_-' + 2 \left(R - r_-\right) r_- \left(3 r_-'' R^2 + r_- \left(\left(r_+'\right)^2 + 3 R r_+'' \right) \right) \Bigr) r_+^2}{\left(R - r_-\right) r_-^2 A_+^2 \left(R - r_+\right)} \\
        &+ \frac{2 \left(r_+ - R\right) A''(r) r_+^2}{A_+^2} 
        - \frac{2 A \left(\omega^2 r_+^3 - \ell (\ell + 1) r_+ + \ell (\ell + 1) R \right) r_+}{R - r_+} 
        + \frac{2 \left(R - r_-\right) A' r_-' r_+}{A_-^2} \\
        &+ \frac{2 A \left(r_- - R\right) A_-' r_-' r_+}{A_-^3} 
        + \frac{2 \left(5 R - 2 r_+\right) A' r_+' r_+}{A_+^2} 
        + \frac{2 A A_+' \Bigl( 3 \left(R - r_+\right) r_+ r_-' + 2 r_- \left(r_+ - 4 R\right) r_+' \Bigr) r_+}{r_- A_+^3} \\
        &+ \frac{2 A A_-' \Bigl( \left(R - r_+\right) r_+ \Bigl( A_+ r_-' + 2 r_- A_+'\Bigr) + r_- A_+ \left(R + r_+\right) r_+'\Bigr) r_+}{A_- r_- A_+^3} \\
        &+ \frac{A \Biggl( \frac{2 \left(R - r_+\right) r_+ \left(A_-'\right)^2}{A_+^2} + r_-' \Bigl( \left(6 - \frac{7 R}{r_-}\right) r_-' + \frac{2 \left(r_- - R\right) r_+'}{R - r_+} \Bigr) \Biggr) r_+}{A_-^2} \\
        &- \frac{A R \Bigl( R r_-' r_+' - 4 \left(R - r_-\right) \Bigl( 3 \left(r_+'\right)^2 + R r_+'' \Bigr) \Bigr) r_+}{\left(R - r_-\right) A_+^2 \left(R - r_+\right)} 
	    - \frac{12 A R^2 \left(r_+'\right)^2}{A_+^2 \left(R - r_+\right)},
	\end{split}
	\end{equation}
	\begin{equation}
		\begin{split} \label{radialC2}
			D_2 =& 
         -\frac{2 A r_-' r_+^4}{r_- A_+^2 \left(R - r_+\right)} 
        - \frac{4 A R \left(r_- - R\right) r_-' r_+^3}{\left(R - r_-\right) r_- A_+^2 \left(R - r_+\right)} 
        - \frac{4 A \left(r_- - R\right) r_+' r_+^3}{\left(R - r_-\right) A_+^2 \left(R - r_+\right)} \\
        &+ \frac{4 \left(r_+ - R\right) A' r_+^2}{A_+^2} 
        + \frac{2 A \left(r_+ - R\right) A_-' r_+^2}{A_- A_+^2} 
        - \frac{2 A R^2 r_-' r_+^2}{r_- A_+^2 \left(R - r_+\right)} 
        + \frac{8 A \left(R - r_+\right) A_+' r_+^2}{A_+^3} \\
        &- \frac{14 A R r_+' r_+^2}{A_+^2 \left(R - r_+\right)} 
        + \frac{2 A \left(R - r_-\right) r_-' r_+}{A_-^2} 
			- \frac{10 A R^2 \left(r_- - R\right) r_+' r_+}{\left(R - r_-\right) A_+^2 \left(R - r_+\right)},  \\[0.5em]
			D_3=& -\frac{2 A r_+^4}{A_+^2 \left(R - r_+\right)} 
        - \frac{4 A R \left(r_- - R\right) r_+^3}{\left(R - r_-\right) A_+^2 \left(R - r_+\right)} 
        - \frac{2 A R^2 r_+^2}{A_+^2 \left(R - r_+\right)}.
		\end{split}
	\end{equation}

\bibliography{BibTeX}

\end{document}